\documentclass[aps,showpacs]{revtex4-1}

\usepackage{amsmath}
\usepackage{amssymb}
\usepackage{graphicx}
\usepackage{dcolumn}
\usepackage{bm}
\usepackage{color}

\newcommand{\be}{\begin{equation}}
\newcommand{\ee}{\end{equation}}
\newcommand{\bea}{\begin{eqnarray}}
\newcommand{\eea}{\end{eqnarray}}
\newcommand{\bma}{\begin{displaymath}}
\newcommand{\ema}{\end{displaymath}}

\begin{document}

\title{Quantum rings for beginners II: Bosons versus fermions}

\author{M. Manninen$^a$, S. Viefers$^{b,*}$, and  S.M. Reimann$^c$}

\affiliation{$^a$Nanoscience Center, Department of Physics, FIN-40014 University of  Jyv\"askyl\"a, Finland \\
                   $^b$ Department of Physics, University of Oslo, P.O. Box 1048 Blindern, 0316 Oslo, Norway \\
                    $^c$Mathematical Physics, LTH, Lund University, P.O. Box 118, S-22100 Lund, Sweden\\
                    $^*${\rm Corresponding author.} Email address: {\rm s.f.viefers@fys.uio.no (S. Viefers)}}

\date{\today}
 
\begin{abstract}
The purpose of this overview article, which can be viewed as a supplement to
our previous review on quantum rings, [S. Viefers {\it et al}, Physica E {\bf 21} (2004), 1-35],
is to highlight the differences of boson and fermion systems 
in one-dimensional (1D) and quasi-one-dimensional (Q1D) quantum rings.
In particular this involves comparing their many-body spectra and other properties, 
in various regimes and models, including spinless and spinful particles, finite versus infinite interaction, 
and continuum versus lattice models. 
Our aim is to present the topic in a comprehensive way, focusing on small systems
where the many-body problem can be solved exactly. Mapping out the similarities and differences between
the bosonic and fermionic cases is of renewed interest due to the experimental developments in recent years,
allowing for more controlled fabrication of both fermionic and bosonic quantum rings.


\end{abstract}


\maketitle


\tableofcontents

\eject

\section{Introduction}

Strictly 1D systems are unusual in that they provide us with one of the rare cases of exactly solvable models for 
a quantum mechanical many-particle system. Thus, a tremendous amount of work has been dedicated to such systems.  
We mention a few selected highlights illustrating the history of the field: 
Tonks\cite{tonks1936} studied the
classical equation of state of elastical spheres in 1D, and the corresponding quantum mechanical
derivation was given by Nagamiya\cite{nagamiya1940}. 
Girardeau\cite{girardeau1960,girardeau1965} showed that 
there is an intimate relationship between impenetrable bosons and fermions in 1D.
Nowadays, 1D gases with infinitely strong short range repulsion thus go by the name
{\it Tonks-Girardeau gases}. 
Lieb and Liniger\cite{lieb1963} as well as Lieb\cite{lieb1963b} and
Young\cite{yang1967} showed that a Bose gas 
with contact interactions has exact 
solutions. (This topic has later on been taken up by many authors, see for
example, Refs.~\cite{sakmann2005,kanamoto2008,cherny2009,ouvry2009,kanamoto2010}).
The Hubbard model for the 1D fermi gas was solved exactly by Lieb and Wu\cite{lieb1968}.
The quantum mechanics in 1D has been reviewed 
by Kolomeisky and Straley\cite{kolomeisky1996} for fermions and 
by Girardeau and Wright\cite{girardeau2002} for Tonks gases.
More recently, Girardeau and Minguzzi\cite{girardeau2007}
have studied quite generally soluble models for 1D gas mixtures, and also
discussed the fermionic Tonks-Girardeau gas~\cite{girardeau2008}. 
Infinitely long 1D fermion systems are usually described as 
Luttinger liquids\cite{luttinger1960,haldane1981}.
In 1D the Fermi surface consists only of two points, which leads to a
so-called Peierls instability\cite{peierls1955} and a breakdown of the Fermi liquid theory.
In such a system the low energy excitations are collective, 
have a phonon-like linear dispersion and thus behave like 
bosons\cite{haldane1994,voit1994,schulz1995}.
Arguably the most exotic property of a Luttinger liquid is {\it spin-charge separation} -- spin and charge
excitations of the interacting 1D system generally move at different velocities.
In fact, in small finite quantum rings where the many-body spectrum can be solved
exactly, the separation of phonon-like charge excitation and spin excitations,
which  can be described in the Heisenberg model, is seen in a very explicit manner\cite{koskinen2001,viefers2004}.

Particles confined in a quasi-one-dimensional ring can support quantum
states exhibiting persistent currents in normal metals\cite{buttiker1983,levy1990}, 
superconductors, quantum
liquids and quantum gases. Persistent currents have mostly been studied in 
normal metals and superconductors (for reviews 
see \cite{eckern1995,viefers2004,kulik2010}).
Superfluidity of helium liquids provides strongly interacting 
atomic systems where persistent currents were 
predicted\cite{reppy1964,grobman1966} and subsequently 
observed for bosons in $^4$He\cite{bendt1962,reppy1964}
and for fermions in $^3$He\cite{gammel1984,pekola1984}.
The possibility of experimentally realizing annular 
traps\cite{guedes1994,nunes1996,felinto1999,hopkins2004,gupta2005,
morizot2006,lesanovsky2007,lesanovsky2007b,heathcote2008,henderson2009} 
for cold atom condensates 
opened up new possibilities to study persistent currents in bosonic  
ring traps\cite{ryu2007,olson2007,henderson2009}.  Such systems have been 
analyzed in many theory works, most of them applying the Gross-Pitaevskii
approximation, but also going beyond; see for example, 
Refs.~\cite{cozzini2006,modugno2006,jackson2006,bao2007,abad2008,ogren2009,malet2010,kaminishi2011,adhikari2012}.  

In metallic or semiconducting quantum rings the current can be
generated by a magnetic flux through a mesoscopic ring.
For neutral atoms confined in a trap,
with electromagnetic fields it is possible to create interactions
which mimic the effect of a magnetic 
flux\cite{mueller2004,heathcote2008,fetter2009,lembessis2010,bruce2011}.
Also, microchip design for persistent current measurements has
been suggested\cite{baker2009}.
In addition, there has been a lot of recent interest in atomic quantum gases 
with dipolar interactions (see~\cite{baranov2008,lahaye2009} for reviews). In ring traps, there 
is an interesting interplay between the anisotropy of the dipolar 
two-body interactions  and the topology of the ring trap; we do not discuss
such dipolar systems any further here, but refer to the recent works in 
Refs.~\cite{abad2010,abad2011,malet2011,zollner2011,karabulut2012}.

The key difference between bosons and fermions is the symmetry
of their many-body wave functions: Exchange of the (spatial and spin) coordinates
switches the sign of the fermion wave function, but leaves the boson wave function unchanged.
In general, the spin of bosons is an integer while that of fermions is a half-integer.
In the case that the Hamiltonian is independent of spin, the spin only adds an
additional degree of freedom which has to be taken into account in requiring
the proper symmetry of the many-body wave function.
It is often illustrative to talk about {\it pseudo-} or {\it iso}spin 
for systems with different components of particles.
We can have one-component fermions, for example, by polarizing 
an electron gas so that each electron is in the spin-up state.
In this case one usually speaks of  {\it ``spinless'' fermions},
in analogy to bosons with spin zero. In the case of 
cold atom clouds one can prepare a bosonic condensate which
has atoms in two different hyperfine states. This is a two-component boson
system, and if we choose to denote the different hyperfine states as different
(pseudo) spin states, we effectively have bosons with pseudospin 1/2.

The purpose of this review
is to highlight the differences of boson and fermion systems 
in one-dimensional (1D) and quasi-one-dimensional (Q1D) quantum rings.
(For harmonic traps set rotating, a comparison between fermion and boson
quantum droplets is given in Refs.~\cite{viefers2008,saarikoski2010}, also discussing the
formation of vortices).

Mapping out the similarities and differences between
bosonic and fermionic rings is of renewed interest due to the experimental progress during recent years,
in particular concerning trapping techniques for cold atoms.
While the original experiments on quantum rings were performed on electronic systems (such as in semiconductor
heterostructures), bosonic as well as fermionic counterparts can nowadays be fabricated with appropriately trapped (charge-neutral) cold atoms.
These systems allow for a much larger degree of control of physical parameters (such as interaction strength) than
is the case for electrons in semiconductor structures. Thus, while a lot of interesting work has been done on Coulomb-interacting electronic rings \cite{loos2012,astra2011,loos2012b,zhu2003}, we here mainly have in mind atomic systems where interactions are typically short-range.

The main emphasis in this paper
is put on small systems where the many-body problem can be solved exactly.
Our intention is to keep the theory as simple as possible, and avoid going into too much detail
of advanced many-particle descriptions. 
The present paper provides a natural supplement to our previous review on quantum rings, Ref.\cite{viefers2004}, 
which we will frequently refer back to.

The outline of the paper is as follows. We first consider one-component boson and fermion systems.
We start with particles interacting with infinitely strong delta function (contact) interactions.
In section \ref{two1inf} we study the exact solution of the two-particle case, and 
in sections \ref{many1inffer} and \ref{many1infbos} we generalize the solution to many particles
and show how the boson and fermion spectra are similar for
odd numbers of particles, but differ for even numbers of particles.
In Section \ref{sec:current} we introduce the concept of persistent currents mentioned previously, and 
demonstrate the relation between the current
and angular momentum in small rings.
In Section \ref{sec:hubbard} we discuss lattice models for rings with only 
one component of particles. 
Section \ref{twocomponent} considers two-component bosons and fermions.
In section \ref{finiteu} we abandon the ideal case of infinitely strong interactions, and study the effect of finite interparticle interactions.
Quasi-one-dimensional systems are considered in Section \ref{quasione}. 

\section{Two spinless particles with infinitely strong contact interactions}
\label{two1inf}

We start with the simplest case, namely two spinless particles confined in an infinitely narrow ring
and interacting with an infinitely strong repulsive contact interaction 
that has the form of a  $\delta$-function, $\delta (x_1-x_2)$. 
When speaking of "spinless fermions", what one typically has in mind is
fully polarized electrons (all in the
same spin state, say spin-up).
The Hamiltonian for such a system is 
(in atomic units $\hbar=m=1$)
\be
H_2=-\frac{1}{2}\left(\frac{\partial^2}{\partial x_1^2}+\frac{\partial^2}{\partial x_2^2}\right)
+g_\infty \delta(x_1-x_2),
\ee
where $x$ is the coordinate along the ring. For convenience we choose $R=1$ and thus
$x$ can be considered as an angle between 0 and $2\pi$. We assume
the limit $g_\infty\rightarrow \infty$, so that the 
wave function is zero whenever $x_1=x_2$ (which for fermions is automatically
satisfied due to the Pauli principle).
With the change of variables $u=x_1-x_2$ and $v=(x_1+x_2)/2$ we get
\be
H_2=-\left(\frac{\partial^2}{\partial u^2}+\frac{1}{4}\frac{\partial^2}{\partial v^2}\right)
+g_\infty \delta(u).
\ee
The solution can be written as
\be
\psi(u,v)=B(u) \sin(n u/2) e^{imv},
\ee
where $B$ is a piecewise constant function needed for bosons and explained below.
The periodic boundary condition $\psi(u,v)=\psi(u,v+2\pi)$ requires 
that $m$ is an integer, and the condition
$\psi(0,v)=\psi(2\pi,v)=0$ requires that $n$ is a positive integer (a negative
integer would only change the sign of the wave function).
Using the original coordinates we can write the wave function as
\be
\psi(x_1,x_2)=B(x_1-x_2)\sin\left(\frac{n}{2}(x_1-x_2)\right)e^{i m (x_1+x_2)/2}.
\ee

We first consider fermions. In this case the function $B$ is not needed (or $B(u)\equiv 1$) since
the rest of the wave function is already antisymmetric with respect to coordinate change.
We have an additional requirement of periodicity
$\psi(x_1,x_2)=\psi(x_1+2\pi,x_2)=\psi(x_1,x_2+2\pi)$,
which is fulfilled only if $n+m$ is even. It is easy to see that in this 
case the solution is a single Slater determinant
\be
\psi(x_1,x_2)\propto \exp\left[i\frac{n+m}{2}x_1\right]\exp\left[i\frac{m-n}{2}x_2\right]
-\exp\left[i\frac{n-m}{2}x_1\right]\exp\left[i\frac{m+n}{2}x_2\right],
\label{2slater}
\ee
consisting of single-particle states with integer angular momenta $(m+n)/2$ and $(m-n)/2$.

The case of bosons is more cumbersome since we have to require that the wave function
is symmetric with respect to particle exchange. This can be done with a 
piecewise constant function $B$ which has constant absolute value but changes sign at 
$x_1-x_2=2\pi k$ where $k$ is an integer. Note that the discontinuity of $B$ is
not a problem since at these points the wave function is zero due to the sine function. 
The required function is then a square wave 
$B(u)={\rm sgn}(\sin(u/2))$, where sgn is the sign function.
The requirement that $\psi(x_1,x_2)=\psi(x_1+2\pi,x_2)$ is then fulfilled for bosons
only if $n+m$ is odd, since adding $2\pi$ either to $x_1$ or to $x_2$ changes the 
sign of $B(u)$. 

Note that the bosonic wave function is not  analytic at $x_1=x_2$ (for odd $n$), 
but has a cusp. However, this is allowed since we assumed that
the contact interaction is infinite and, consequently, the wave function
has a node at that point.

In general, we can now write the eigenvalues for the two-particle system 
with infinitely strong  contact interaction as 
\be
\epsilon_{n,m}=\frac{1}{4}(n^2+m^2),
\ee
where $n+m$ has to be even for fermions and odd for bosons.

\begin{figure}[h!]
\includegraphics[width=.5\columnwidth]{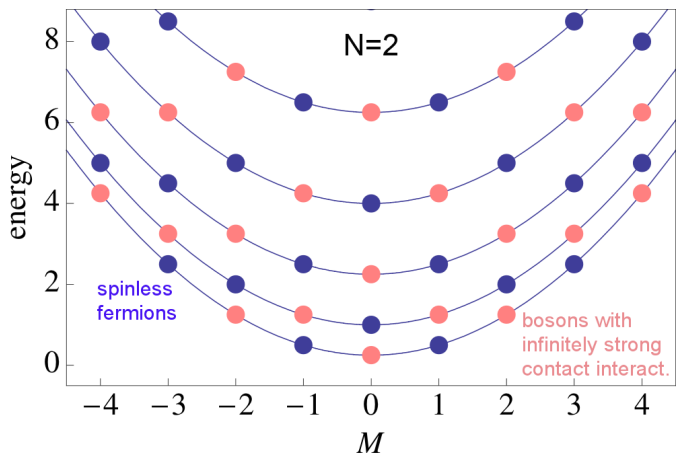}
\caption{Energy spectra for two spinless fermions (blue) and for two
bosons interacting with infinitely strong delta function interactions (pink).
The lines show the different 'vibrational modes'
}
\label{bf2}
\end{figure}

The total angular momentum of the state is $m$, since
\be
\hat L\psi=-i\left(\frac{\partial}{\partial x_1}+\frac{\partial}{\partial x_2}\right)\psi
=-i\frac{\partial}{\partial v}\psi=m\psi.
\ee
The wave function thus consists of an internal part, $\sin(nu/2)$, and a center of mass part,
$\exp(imv)$. The latter corresponds to a rigid rotation of the two-particle
system,  while
the internal part corresponds to an internal vibrational mode\cite{viefers2004}.
Note that due to symmetry restrictions, we get the same selection rules 
as for two-atomic molecules:
Purely rotational states ($n=1$) are allowed for fermions only with odd angular momenta,
and for bosons only for even angular momenta\cite{tinkham1964}. 
Figure \ref{bf2} shows the energy spectra displaying the alternating allowed fermion and
boson states for each of the vibrational states.

The ground state of two bosons ($m=0, n=1$) has zero angular momentum, energy $E=1/4$ and, interestingly,
it naively looks like a function of single-particle states with {\it half-integer} angular momenta 
(or half of the mass),
\bea
\Psi_{\rm GS}^{\rm 2 bosons}
&=&{\rm sgn}\left[ \sin \left( \frac{x_1-x_2}{2} \right) \right] \sin\left(\frac{x_1-x_2}{2}\right) \nonumber \\
&=&{\rm sgn}\left[ \sin \left( \frac{x_1-x_2}{2} \right) \right] \frac{1}{2i}\left[e^{ix_1/2}e^{-ix_2/2}-e^{-ix_1/2}e^{ix_2/2}\right].   \label{2bgs}
\eea
These, however, are not single-particle eigenstates of the ring.
Fourier expansion in terms of the true single-particle eigenstates gives an infinite Fourier series
(denoting, as earlier, $x_1-x_2=u$)
\bea
\Psi_{\rm GS}^{\rm 2 bosons}&=&
\frac{2}{\pi}+\sum_{m=1}^\infty \frac{4\cos(mu)}{\pi(1-4m^2)} \\
&=&
\frac{2}{\pi}+\sum_{m=0}^\infty \frac{2}{\pi(1-4m^2)}\left[e^{imx_1}e^{-imx_2}+e^{-imx_1}e^{imx_2} \right].
\label{2bosexp}
\eea
In the case of spinless fermions the antisymmetry dictates that a single Slater determinant is 
always a solution for contact interactions, as pointed out above. The ground state, found by setting $n=m=1$ in (\ref{2slater}), equals
$\exp(ix_1)-\exp(ix_2)$. It
has angular momentum 1 and energy $E=1/2$, while the state with zero angular momentum ($m=0, n=2$) is
\bma
\Psi_{L=0}^{\rm 2 fermions}=
e^{ix_1}e^{-ix_2}-e^{-ix_1}e^{ix_2} 
\ema
and has higher energy $E=1$. Both energies are higher than the ground
state energy for bosons ($E=1/4$), as illustrated in Fig. \ref{bf2}.

In contrast to spinless fermions, for bosons the interaction has a true effect.
This allows the bosonic wave function to 'take advantage' of all single-particle states, 
like in Eq. (\ref{2bosexp}), for lowering the energy.

\section{Spinless fermions with contact interactions (noninteracting fermions)}
\label{many1inffer}

In this section we discuss the so-called {\it yrast} states, {\it i.e.} lowest energy states for given total angular momentum, 
of non-interacting spinless fermions in a strictly 1D ring. Most of the detailed analysis is presented for an odd number of particles,
with the corresponding results for even $N$ stated at the end.

The ground state of $N$ noninteracting fermions is a Slater determinant of the lowest
$N$ single-particle states. In a strictly one-dimensional (1D) ring
with radius $R=1$ the single-particle energies and wave functions are
\be
\epsilon_\ell=\frac{\ell^2}{2},\qquad \phi(x)=e^{i\ell x},
\label{spenwf}
\ee
where the $\ell$ is the single-particle angular momentum and $x$ the coordinate
along the ring (equal to the angle in polar coordinates).
In the ground state the $N$ lowest states are occupied.
For an odd $N$  the single-particle states from $\ell=0$ to
$\ell=\pm (N-1)/2$ are occupied and the energy becomes 
\be
E_{\rm gs}^{\rm ODD}(N)=2\sum_{\ell=1}^{(N-1)/2} \frac{\ell^2}{2}=\frac{1}{24}N(N^2-1).
\label{gsODDN}
\ee
The angular momentum of the ground state is zero.
The corresponding ground state wave function for an odd number $N$ of fermions is the Slater determinant
\bma
\Psi_{\rm GS}(x_1,x_2,\cdots,x_N)=
\left|\begin{array}{cccc}
e^{-inx_1}&e^{-inx_2}&\cdots&e^{-inx_N}\\
e^{-i(n-1)x_1}&e^{-i(n-1)x_2}&\cdots&e^{-i(n-1)x_N}\\
\cdots&\cdots&\cdots&\cdots\\
e^{inx_1}&e^{inx_2}&\cdots&e^{inx_N}\\
\end{array}\right|,
\ema
where $n=(N-1)/2$. Using the mathematical properties of a determinant we can recast this as
\bma
\Psi_{\rm GS}=e^{-inx_1}e^{-inx_2}\cdots e^{-inx_N}
\left|\begin{array}{cccc}
1&1&\cdots&1\\
e^{ix_1}&e^{ix_2}&\cdots&e^{ix_N}\\
\cdots&\cdots&\cdots&\cdots\\
e^{i2nx_1}&e^{i2nx_2}&\cdots&e^{i2nx_N}\\
\end{array}\right|
\ema
\bma
=e^{-inx_1}e^{-inx_2}\cdots e^{-inx_N}\prod_{i<j}^N(e^{ix_i}-e^{ix_j}),
\ema
where in the last step we have identified the determinant as a so-called Vandermonde determinant \cite{gradshteyn1980}.

By noticing that each $x_i$ appears in $2n$ factors of the product ($2n=N-1$),
we can distribute the exponentials in front into these factors and get  \cite{mitas2006}
\bea
\Psi_{\rm GS}=\prod_{i<j}^N (e^{i(x_i-x_j)/2}-e^{i(x_j-x_i)/2})= (2i)^{N(N-1)/2}.
\prod_{i<j}^N \sin\left(\frac{x_i-x_j}{2}\right).
\label{eq:gsodd}
\eea
Since we are working with unnormalized wave functions, the overall constant factor will be ignored in the following.

It is easy to verify that multiplying the wave function of this (or any other) many-particle eigenstate with the symmetric 
function $\exp(i \sum x_i)$ will again generate a solution of the many-particle problem, with an energy increase $E \rightarrow E + M + N/2$ where $M$ is the total angular momentum of the original state.
This operation corresponds to an increase of the angular momentum of each 
single-particle state by one.
Exciting the ground state this way (recalling that its total angular momentum is zero) will thus increase the angular momentum by $\Delta M=N$, and the total energy 
by $N/2=\Delta M^2/2N=\Delta M^2/2I$. 
This term in fact has the same form as 
the classical rotational energy of a rigid body, with moment of inertia 
$N$ (we have chosen the radius and mass to be equal to one). For this reason we will
refer to it, somewhat sloppily, as "classical rotation energy", although of course it results 
from a purely quantum mechanical derivation. 
More generally, we find that the lowest energy states for total angular momenta $M=mN$
are obtained through rigid rotation of the ground state, with wave function
\bma
\Psi_{[M=mN]}=\prod_{i}^N e^{imx_i} \prod_{i<j}^N \sin\left(\frac{x_i-x_j}{2}\right),
\ema
and corresponding energy
\be
E_{\rm L=mN}^{\rm ODD}(N)=\frac{1}{24}N(N^2-1)+\frac{M^2}{2N},
\label{erot}
\ee 
where the last term is again recognized as corresponding to rigid rotation of the entire ring.

For angular momenta different from $mN$ we can also easily determine the lowest
energy from the single Slater determinant of each state. For example, the lowest state
for $M=1$ is obtained by exciting the topmost single-particle from $\ell=(N-1)/2$ to
$\ell=(N+1)/2$. In terms of wave functions this means multiplying the 
ground state with a symmetric polynomial of exponentials $\exp(ix_j)$, 
\bea
\Psi_{[M=1]}=\left(\sum_{i}^N e^{ix_i}\right) \prod_{i<j}^N \sin\left(\frac{x_i-x_j}{2}\right),
\eea
with energy $E_{M=1}=E_{\rm GS}+N/2$. We note that, curiously,  the energy is the same for
$M=1$ and $M=N$.
This energy increase, $N/2$, is much
larger than what the classical rotational energy, $1/2N$, would be. 
The energy difference is a {\it vibrational energy} of the quantum state.

It is straightforward to calculate the lowest vibrational energy for any 
angular momentum $\nu$ ($1 \le \nu \le M$), i.e. the energy difference between the 
true quantum mechanical energy and the 'classical' rotational energy:
\be
\Delta E_{\rm v}(\nu) =\frac{\nu(N+1-\nu)}{2} - \frac{\nu^2}{2N},
\label{evib}
\ee
where the first term is the lowest excitation energy from the $M=0$ to the $M=\nu$ state,
obtained by exciting the particle at $\ell=(N-1)/2 - (\nu-1)$ to $\ell=(N-1)/2 +1$ (see Fig.1 in Ref.\cite{viefers2004}),
and the second term is the subtraction of the classical rotation energy.
Since any state can be multiplied by  the
rigid rotation $\exp(i m\sum x_i)$, we obtain that the rotational spectrum 
has 'a period of $N$' in the sense that the {\it internal structure}
of the state does not change when the angular momentum is 
increased by $N$ \cite{viefers2004}. This means that the vibrational energy 
must be the same for $M=\nu$ and $M=\nu+mN$ ($m$ integer).
The above 'phonon' spectrum, taking the form of an inverted parabola, is analogous to
a classical row of particles interacting with $1/d^2$ 
interaction\cite{calogero1969,sutherland1971,calogero1971,viefers2004}.
In quantum mechanics the effective $1/d^2$ interaction results
from the kinetic energy: A particle confined between the neighboring
particles is essentially a particle in an infinite box with kinetic energy 
proportional to $1/d^2$. 
Classical particles interacting only with a contact interaction would
not have vibrational states.

We can now construct the lowest energy for each angular momentum, i.e.
the yrast states, by adding the ground state energy for $M=0$,
i.e. Eq. (\ref{gsODDN}), the 'classical' rotational energy $M^2/2N$, 
and the vibrational energy of Eq. ({\ref{evib}), taking into account that it
is the same for $\nu$ and $\nu+mN$. The result can be written as
\be
E_{\rm yrast}^{\rm ODD}(N,M)=
\frac{1}{2}(N+1)(M \bmod N)-\frac{1}{2}(1+\frac{1}{N})(M \bmod N)^2+\frac{M^2}{2N}
+\frac{1}{24}N(N^2-1),
\label{oddenergy}
\ee
as may be independently verified by a numerical computation of the quantum mechanical spectrum. 
Higher 'vibrational' excited states can be derived from the single-particle energies in a similar manner.

Finally, let us state the corresponding main results for an {\it even} number of spinless fermions.
In this case the ground state can again be expressed as a Slater 
determinant, but now the angular momentum is not zero but rather $M=\pm N/2$.
Manipulations similar as those leading to Eq.(\ref{eq:gsodd}) give the wave functions for the two degenerate ground states
\be
\Psi_{\rm GS}^{\rm fermions}=
\prod_{i}^N e^{\pm ix_i/2} \prod_{i<j}^N \sin\left(\frac{x_i-x_j}{2}\right),
\label{efwf}
\ee
where total angular momentum can be immediately read off from the first product.
An analysis similar to the one leading to Eq.(\ref{oddenergy}) can be carried out to find the yrast energies,
\be
E_{\rm yrast}^{\rm EVEN}(N,M)=\frac{1}{2}(N+1)((M+N/2) \bmod N)-
\frac{1}{2}(1+\frac{1}{N})((M+N/2) \bmod N)^2+\frac{M^2}{2N}
+\frac{1}{24}N(N^2-1).    
\ee

\section{Bosons with infinitely strong  contact interactions}
\label{many1infbos}

For odd numbers of bosons the solutions of energy 
levels are equal to those of spinless fermions.
This can be most easily shown in a lattice model which we will
return to in section \ref{sec:hubbard}. 
The ground state wave function can be obtained from the 
fermion solution by multiplying it with an appropriate product of sign functions,
to make the state symmetric with
respect to particle coordinate change.
Like in the two-particle case, the boson
wave function is not analytic, since at the points $x_i=x_j$ the partial derivatives
are not continuous. However, this is not required since 
we assume an infinitely strong repulsive interaction (in fact even a finite
contact interaction causes a discontinuity in the derivative of the wave function\cite{lieb1963}).
For example, the ground state for an {even} number of bosons is a direct generalization of the two-particle case (\ref{2bgs}),  
\bma
\Psi_{\rm GS}^{\rm bosons}=
\prod_{i<j}^N {\rm sgn}\left[ \sin\left(\frac{x_i-x_j}{2}\right) \right]\sin\left(\frac{x_i-x_j}{2}\right),
\ema
and has angular momentum $M=0$.

\begin{figure}[h!]
\includegraphics[width=.5\columnwidth]{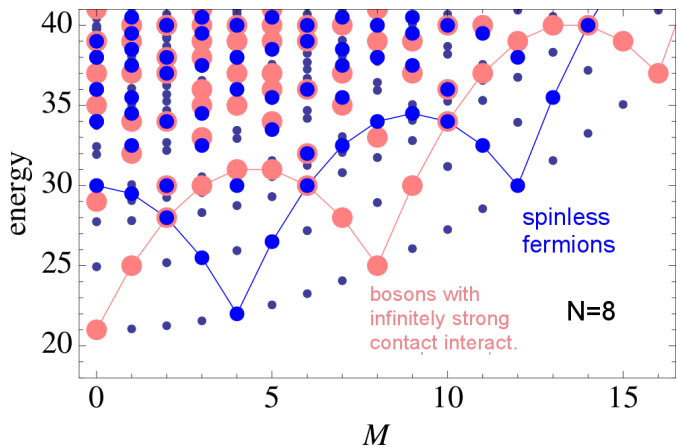}
\caption{Energy spectra for 8 spinless fermions (blue) and for 8
bosons interacting with infinitely strong contact interactions (pink)
in a 1D quantum ring of radius $R=1$.
The small dots show forbidden states. The blue and pink lines show the 
yrast energies for fermions and bosons.
}
\label{ener8}
\end{figure}

In the boson case the energy of the yrast line was recently studied in Ref.\cite{kaminishi2011}.
It is qualitatively similar for odd and
even numbers of particles and is expressed in both cases by Eq. (\ref{oddenergy}).
This means that for odd numbers of particles the yrast lines for bosons and fermions are identical to each other,
while for even numbers they differ qualitatively.
Figure \ref{ener8} shows the energy spectra for 8 spinless fermions and bosons
with infinitely strong contact interactions. Like in the case of two particles,
the boson and fermion states have different 'selection rules' which allow
combination of the angular momenta of the 'vibrational states'.

Again we find that the energy spectrum is periodic in $N$ in 
the following sense: For any energy eigenvalue $E_{[M]}$
we have $E_{[M^\prime=M+kN]}=E_{[M]}+k^2N/2$, 
and at the yrast line $E_{[M=1]}=E_{[M=N]}=E_{\rm GS}+N/2$.
In the case of bosons this holds both for odd and even numbers of particles.

As already mentioned in the introduction, we here stress again that 
boson systems with finite contact interaction 
have been solved exactly already long time ago by
Lieb and Liniger\cite{lieb1963,lieb1963b} and \cite{yang1967}.
and later on many others (see for example,
Refs.~\cite{sakmann2005,bao2007,kanamoto2008,cherny2009,ouvry2009,
kanamoto2010,kaminishi2011}, this list being
far from complete).

\section{Persistent current and angular momentum}
\label{sec:current}

In the case of electrons, or other {\it charged} particles, 
a persistent current can be introduced by a magnetic flux through the ring.
Neutral particles obviously do not respond to magnetic fields in the same way.
However, as mentioned in the Introduction, there are ways of mimicking the effect of a magnetic 
flux on, e.g. charge neutral bosonic atoms \cite{mueller2004,heathcote2008,fetter2009,lembessis2010,bruce2011}. 
Thus, the following discussion is relevant also to such systems, although we will stick to the standard terminology of 'real' magnetic fields. 

The physics of persistent currents becomes particularly transparent if one chooses the magnetic field at the ring radius to zero, while the 
flux through the ring is nonzero. This can be modeled by an appropriate choice of vector potential\cite{viefers2004}. In such a system the single-particle wave functions of the 
ideal 1D ring do not change due to the flux, while the single-particle energies are modified (for a derivation, see e.g. \cite{viefers2004}), 
\be
\epsilon_m(\Phi)=\frac{1}{2}\left(m-\frac{\Phi}{\Phi_0}\right)^2,
\label{speflux}
\ee 
where $\Phi$ is the flux through the ring and $\Phi_0$ the flux quantum $\Phi_0=h/e$
($\Phi_0=2\pi$ in our units where $\hbar =e=1$). 
The many-particle state for noninteracting electrons (or spinless electrons with
contact interactions) is still the same single Slater determinant as without
the flux, but the total many-particle energy becomes
\be
E_M(\Phi)=E_M(0)-M\left(\frac{\Phi}{\Phi_0}\right)+\frac{N}{2}\left(\frac{\Phi}{\Phi_0}\right)^2.
\label{mbeflux}
\ee
In the case of bosons interacting with an infinitely strong contact 
interaction, 
the many-particle state consists of many permanents made of the single-particle 
wave functions. Since the flux does not change the single-particle states and each
permanent must have the same angular momentum, the flux does not change the
many-body wave functions. The many-particle energy for bosons will also be 
described with Eq. (\ref{mbeflux}) above.

\begin{figure}[h!]
\includegraphics[width=.5\columnwidth]{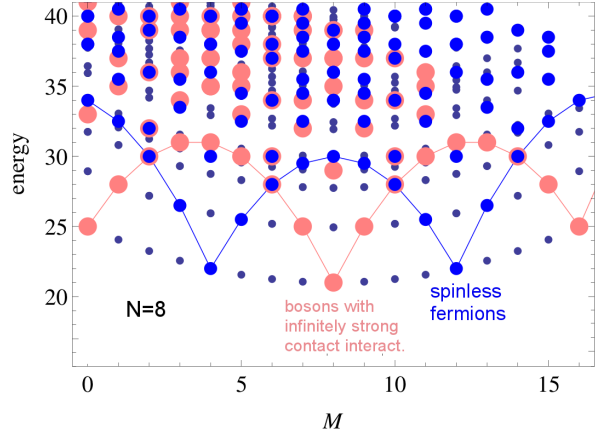}
\caption{Energy spectra for 8 spinless fermions (blue) and for 8
bosons interacting with infinitely strong contact interactions (pink)
in a 1D quantum ring of radius $R=1$. A magnetic flux $\Phi=\Phi_0$
passes through the ring.
The small dots show forbidden states. The blue and pink lines show the 
yrast energies for fermions and bosons.
}
\label{ener8flux}
\end{figure}

The current (density) for a single-particle state $\phi_\ell$ in a 1D system, with a flux $\Phi$ passing through
the ring (of radius $R=1$ and in units $\hbar=e=1$), is\cite{viefers2004}
\be
j_\ell=\frac{-i}{4\pi}\left(\phi_\ell^*(\frac{\partial}{\partial x}-i\frac{\Phi}{\Phi_0})\phi_\ell(x)-
\phi_\ell(\frac{\partial}{\partial x}+i\frac{\Phi}{\Phi_0})\phi_\ell^*(x)\right)
=\frac{\ell}{\Phi_0}-\frac{\Phi}{\Phi_0^2}.
\label{spcurrent}
\ee

It follows that the current of the many-particle state, consisting of Slater determinants 
or permanents, will be \be
J=M/\Phi_0 - N\Phi/\Phi_0^2.
\label{mbcurrent}
\ee
This is because each determinant (permanent) has the same total angular momentum $M$ and
consists of $N$ single-particle states.

Generally, it can be shown that the current caused by the flux can be derived from the
flux dependence of the total energy as\cite{byers1961,viefers2004} 
\be
J=-\frac{\partial E}{\partial \Phi}.
\label{current}
\ee
Clearly, when applied to Eq. (\ref{mbeflux}) the above equation gives Eq. (\ref{mbcurrent}).
At finite temperatures the current can be derived by replacing the total energy $E$
with the Helmholtz free energy $F$ in Eq. (\ref{current}).

\begin{figure}[h!]
\includegraphics[width=0.8\columnwidth]{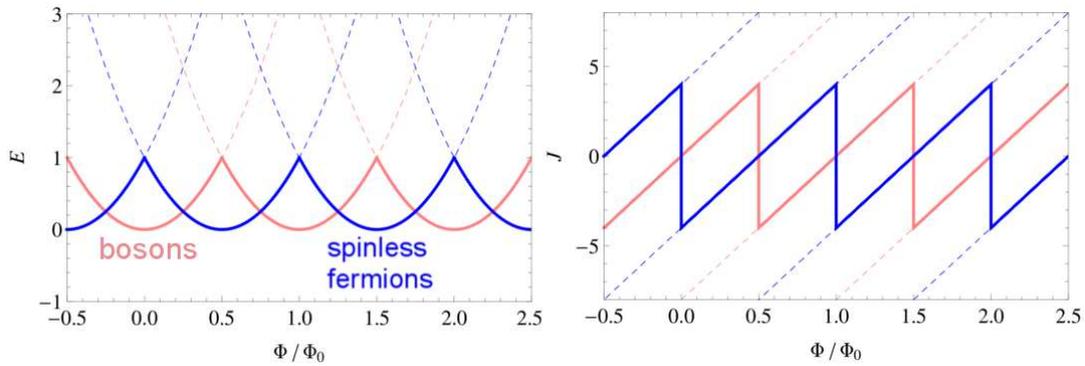}
\caption{Energy spectra for 8 spinless fermions (blue) and for 8
bosons (pink) as a function of the flux passing through the ring (left), and the 
corresponding persistent current (right). Energies and currents corresponding 
to different angular momenta are shown as dashed lines and those
corresponding to the ground state as solid lines.
The particles interact with an infinitely strong contact interaction.
}
\label{current8}
\end{figure}

Comparing figures \ref{ener8} and \ref{ener8flux} we notice that the effect of the flux
is to tilt the spectrum  so that energy states with higher angular momenta become
the lowest energy states. When we determine the persistent current for the ground state
we notice that we jump from one angular momentum state to another when the flux increases.
For 8 bosons these states are 0, 8, 16 etc., and for fermions 4, 12, 20 etc. 
(see left panel in Fig.~\ref{current8}).
The current 
as a function of the flux for 8 particles interacting with infinitely strong 
contact 
interaction is shown in the right panel of Fig. \ref{current8}.

\section{Hubbard model with infinite interaction ($t$-model)}
\label{sec:hubbard}

\subsection{Tight-binding basis and spinless particles}

The Hubbard model\cite{hubbard1963} has been extensively used for studying 
persistent currents in quantum rings (see e.g.\cite{fye1991,yu1992,kusmartsev1995,viefers2004}).
The basis of the Hubbard model is the tight binding (TB) model: The electrons are 
predominantly localized on atoms at lattice sites, but are allowed to hop between neighbouring
sites. In a 1D lattice the simple TB model is, somewhat surprisingly, 
also closely related to the free electron model.
A numerical solution of the free electron Hamiltonian in a lattice leads to the
simple TB model with one state per lattice site\cite{manninen1991}. 
This means that the low energy
spectrum of the TB model is identical with that of the free-electron case.
The simple TB Hamiltonian for the ring is
\be
H=-t\sum_{i=1}^L a_i^\dagger a_{i+1} +{\rm HC},
\label{tbhamiltonian}
\ee
where $a_i^\dagger$ and $a_i$ are creation and annihilation operators for site $i$,
HC denotes Hermitian conjugate, and the periodic boundary condition requires that 
$a_{L+1}^\dagger=a_1^\dagger$.
The corresponding Hamiltonian matrix has nonzero matrix elements 
$H_{ij}=-t$ only if $i$ and $j$ are nearest neighbours. 
The eigenvalues for the Hamiltonian matrix for a ring of $L$ sites
and radius $R=1$ are (assuming $L>2$)
\be
\epsilon(n)=-2t\cos\left(\frac{2\pi}{L}n\right),
\label{tbse}
\ee
and the corresponding eigenvectors (wave functions at site $j$)
\be
\phi_n(j)=e^{i 2\pi n j/L}.
\label{tbsv}
\ee
From the periodic boundary condition it follows that 
$\epsilon(n)=\epsilon(L-n)=\epsilon(-n)$ and
$\phi_n(j)=\phi_{L-n}(-j)=\phi_{-n}(-j)$.
We notice that in a large ring, with $L\gg n$ the low energy eigenvalues 
are those of free electrons, $\epsilon(n)\approx -2t+t k^2$,
with the wave vector $k=2\pi n/L$.

We will first consider the many-body problem of spinless particles interacting with
an infinitely strong contact interaction in quantum rings made of $L$ lattice sites. 
The infinitely strong repulsive short strange
interaction prevents two or more particles from occupying the same lattice site. 
The many-body Hamiltonian describing such a system is the Hubbard model
\be
H=-t\sum_{ij}^{\rm nn} a_i^\dagger a_{j} +\frac{U}{2}\sum_i \hat n_i(\hat n_i-1),
\label{hubbard}
\ee
where "nn" means that the sum goes only over nearest neighbors,
$\hat n_i=a_i^\dagger a_i$ and $U$ the onsite interaction. 
Note that for spinless fermions the interaction term vanishes since $\hat n_i^2=\hat n_i$.
Also in the case of bosons with $U\rightarrow\infty$ we can neglect the
interaction term of the Hamiltonian 
if we restrict the many-particle basis to states with only zero or single occupancy in each site. 
The Hamiltonian then reduces to the tight binding Hamiltonian, Eq. (\ref{tbhamiltonian}).
This model is sometimes called the $t$-model.

We describe the many-body states in the localized basis and write the basis vectors as
$\vert n_1n_2n_3\cdots n_L\rangle$, where occupation numbers $n_i$ are now
restricted to be 0 or 1 (also for bosons) and the total number of particles is $\sum_i n_i=N$.
The only difference between the bosons and fermions is thus whether 
commutation or anticommutation relations are used for the operators $a$ and 
$a^\dagger$. Consider one term of the Hamiltonian, $a_i^\dagger a_{i+1}$, where $i<L$ ,
operating on a state with $n_{i+1}=1$ and $n_i=0$. Both for fermions and bosons 
we get 
\be
a_i^\dagger a_{i+1}\vert \cdots 01  \cdots\rangle=
+\vert \cdots 10  \cdots\rangle,
\ee
independent of the occupation of the other states. This is because
the creation and annihilation operators operate on neighbouring positions 
and, consequently, 
the phase factors for fermion operators (powers of (-1)) always add to an even number.
A difference can appear when $i=L$, since $a_{L}^\dagger a_{L+1}=a_{L}^\dagger a_1$.
In this case the operation with $a_{L+1}=a_1$ will give no phase factor but the 
subsequent operation with $a_L^\dagger$ gives a phase $(-1)^{N-1}$ for fermions.
Clearly this is $+1$ if $N$ is odd and $-1$ if $N$ is even.
The periodic boundary conditions thus give
\be
a_L^\dagger a_{1}\vert 1 \cdots 0\rangle=
\begin{cases}
-\vert 0\cdots1\rangle&\text{for even number of fermions} \\
+\vert 0\cdots1\rangle&\text{otherwise}. \\
\end{cases}
\ee
The important result is that if the number of particles is odd, 
the Hamiltonian matrices for bosons and fermions are identical, 
leading to the same eigenvalues and eigenvectors. 
However, two things have to be noticed: (i) the eigenvectors 
are the same only if the localized single-particle basis is used 
for constructing the many-body states and (ii) the meaning of the
many-body basis vector $\vert n_1n_2n_3\cdots n_L\rangle$ is different
for bosons and fermions.
In the case of fermions it is an antisymmetric Slater determinant 
while in the case of bosons it is a symmetric function called permanent.

Remembering that at low energies the TB model is equal to the free
particle model, it is easy now to reconsider the particles in a continuous
ring studied in sections \ref{many1inffer} and \ref{many1infbos}.
In the case of odd numbers of particles the energy spectra of bosons and fermions are identical,
while for even numbers of particles, as shown in Figs. \ref{bf2} and
\ref{ener8}, they are still related, but qualitatively different.

\begin{figure}[h!]
\includegraphics[width=.5\columnwidth]{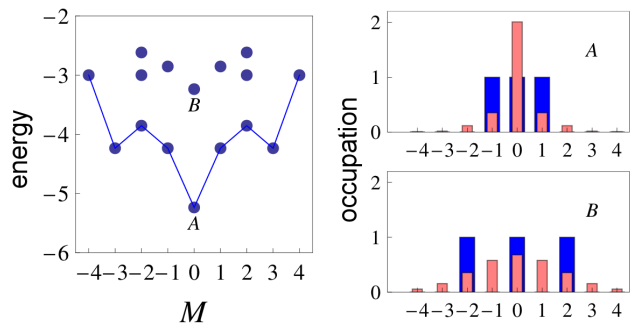}
\caption{Energy spectrum of three spinless particles in a Hubbard ring with 
10 sites and infinite $U$. The left panel shows the lowest energy levels (dots) as a function
of the angular momentum $M$. The yrast levels are connected with a line to
guide the eye.
The spectrum is identical for bosons and fermions.
The right panels show the occupation of the TB single-particle states of the ring
for the ground state (A) and for an excited state (B), blue for fermions and 
pink for bosons. The fermion states are single determinants with only three
of the states occupied, while the boson states consist of numerous 
permanents.
}
\label{bf3}
\end{figure}

We could also use the single-particle states of the TB model, Eq. (\ref{tbsv}),
when constructing the many-particle basis for bosons and fermions.
In this case all fermion wave functions would be single Slater determinants,
as we noted above (the contact interaction has no meaning for
 spinless fermions).
However, a typical boson state would be a complicated linear combination of several 
permanents.
This difference is illustrated in Figure \ref{bf3} which 
shows the energy spectrum for three particles in a
ring of ten sites ($N=3$, $L=10$). The spectrum is identical for bosons and fermions.
The state vectors are also identical in the occupation number representation
when the localized basis is used. However, in the TB basis, where the fermion
wave functions reduce to single Slater determinants, the boson wave functions remain
complicated as shown in Fig. \ref{bf3} where the occupations of
the TB single-particle states are shown for the ground state and for an excited state.
In the case of fermions only three TB states are occupied. In the case of bosons 
all single-particle states have some occupation.
Note that the infinitely strong contact interaction leads to a significant
depletion of the single particle ground state, which in the non-interacting case would be occupied by all $N$ bosons..

The rotational spectrum of a Hubbard ring is characterized
by a periodicity of $L$. In the case when $N\ll L$ the yrast line also shows a
periodicity of $N$ in a similar way as in the case of continuous rings. This is illustrated
in Fig. \ref{bf3}, showing the yrast line for $N=3$ which has minima at angular momenta -3, 0, and 3.
Lattice quantum rings thus have all the same features as continuous rings when
particles with strong contact interaction are considered.

\subsection{Magnetic flux and current in a lattice}

The effect of a magnetic flux piercing a ring of $L$ lattice sites is to add
a phase in the hopping term of the TB Hamiltonian\cite{peierls1933,viefers2004}
\be
H=-t\sum_i^L\left(e^{-i2\pi\Phi/(\Phi_0L)} a_i^\dagger a_{i+1}
+e^{i2\pi\Phi/(\Phi_0L)} a_{i+1}^\dagger a_i\right).
\label{tbfluxh}
\ee
In the lattice we can define the current operator between two
points $i$ and $j$ generally as
\be
J_{ij}=-i \frac{2\pi}{\Phi_0 L} \left( t_{ij} e^{i \Phi_{ij}}a_i^\dagger a_j - t_{ji} e^{i \Phi_{ji}}a_j^\dagger a_i\right) ,
\label{tbcurrent}
\ee
where $t_{ij}$ is the hopping parameter and $\Phi_{ij}$ the phase change
between the two points. In a simple ring $\Phi_{i,i+1}=2\pi\Phi/(\Phi_0L) = -\Phi_{i+1,i}$,
and the current is the same between any neighbouring sites.
Alternatively, we can again determine the current as the derivative
of the total energy, using Eq. (\ref{current}).
Note that when the ring is quasi-one-dimensional, Eq. (\ref{tbcurrent}) is 
still valid, but for calculating the total current one has to sum the currents 
through different paths.

\subsection{Particle-hole symmetry}

In a Hubbard ring the persistent current can be determined as the derivative of the 
total energy as a function of the flux (Eq. (\ref{current})
or by computing the expectation value
of the current operator between two adjacent points, Eq. (\ref{tbcurrent}).

In the case of bosons with infinitely strong contact interaction
($U\rightarrow \infty$) the Hamiltonian is the same for particles and holes, 
i.e. $H=\sum a_i^\dagger a_j =\sum a_j a_i^\dagger$ since
the operators commute when $i\ne j$. This means that
the ground state energy and the 
persistent current are symmetric with respect to 
particles and holes, irrespective of the symmetry of the 
single-particle spectrum. The situation is different for 
fermions due to the anticommutation rule, which changes
the sign of the Hamiltonian for holes. Consequently,
in the case of fermions the many-body energy and 
the persistent current of the ring are symmetric with
respect to particles and holes only if the single-particle spectrum is symmetric.

\section{Two-component systems (spin-1/2)}
\label{twocomponent}

\subsection{Hubbard model with infinite $U$}
\label{secinfu}

Let us now now consider particles with spin (e.g. normal spin for fermions and two 
different hyperfine states for bosons), but still apply  the same simple 
Hubbard model with $U\rightarrow \infty$.
The spin increases the size of the Fock space since all spin configurations should be included.
Since we still assume the infinite $U$-, or $t$-model, only one particle is allowed
in a lattice site, independent of its spin. 
The energy spectrum for bosons and fermions is the same if both 
of the spin components have an odd number of particles, or if the total number of
particles is odd.
In the former case we can arrange the single-particle states as
$\vert n_{1\uparrow},n_{2\uparrow},\cdots,n_{L\uparrow};
n_{1\downarrow},\cdots n_{L\downarrow}\rangle$. Acting with
the fermion operator $a_{i,\sigma}^\dagger a_{i+1,\sigma}$ does not change the sign.
In the latter case we arrange the basis states as
$\vert n_{1\uparrow},n_{1\downarrow},n_{2\uparrow},
n_{2\downarrow},\dots,n_{L\uparrow},
n_{L\downarrow}\rangle$, and again 
$a_{i,\sigma}^\dagger a_{i+1,\sigma}$ can not change the sign (note that only
one of $n_{i+1,\uparrow}$ and $n_{i+1,\downarrow}$ can be nonzero
since only one particle is allowed in any site.

The case when both $N_\uparrow$ and $N_\downarrow$ are even, is 
more complicated. However, also in this case all the eigenvalues 
are still the same for bosons and fermions, but the degeneracies 
of the eigenvalues can differ. 

In general, for a fixed total number of particles, $N=N_\uparrow+N_\downarrow$,
all the energies of the spectrum are the same for any combination of nonzero
$N_\uparrow$ and $N_\downarrow$, but the degeneracies can differ.
To prove this, we first notice that a state with $N_\uparrow=N_\downarrow$
(or $N_\uparrow=N_\downarrow+1$ for odd number of particles)
has $S_z=0$ (or $S_z=1/2$), where $S_z=N_\uparrow-N_\downarrow$ is the $z$-component of the total spin. 
Since the Hamiltonian is independent
of the spin, the spectrum of $S_z=0$ includes all energies for states with
$S_z=-N,~-N+1,\cdots,~N$, i.e. for all
possible values of  $S_z=N_\uparrow-N_\downarrow$.
We then only have to show that $N_\uparrow =N-1$ and $N_\downarrow=1$ already 
gives all possible eigenvalues. This can be done with the help of the 
Bethe ansatz solution, as will be discussed in the case of an 
infinitely strong contact interaction in the section below.

It is interesting to observe, that
adding to a one-component system just one particle of the other
component will change the energy spectrum completely by 
allowing now {\it all}  the states, which were forbidden by the wave function 
symmetry in the one-component case.
For example, in the eight particle case of Fig. \ref{ener8} all the energies
shown with black dots would become allowed for both bosons and fermions
if the 8 particles consisted of, say, 7 "spin-up" particles and one "spin-down" particle.

\subsection{Bethe ansatz solution for particles with infinitely strong contact 
 interactions}

The Hubbard model for fermions in a 1D lattice was solved exactly 
by Lieb and Wu\cite{lieb1968} using the Bethe ansatz. The total energy 
of a given many-body state with $N$ particles can be written as
\be
E=-2t \sum_{j=1}^N \cos{k_j},
\label{betheenergy}
\ee
where the numbers $k_j$ are found by solving a set of algebraic (transcendental)
Bethe equations (see e.g. \cite{viefers2004}). In the limit of infinite $U$ these numbers
will be
\be
k_j=\frac{2\pi}{L}\left[ I_j-\frac{p}{N}\right],
\ee
with 
\be
p=-\sum_{\alpha=1}^{N_\uparrow} J_\alpha.
\ee
The quantum numbers $I_j$ and $J_\alpha$ are related to the charge and spin
degrees of freedom, respectively. Their possible values depend on whether 
$N$ and $N_\uparrow$ are even or odd. Here we will only consider even $N$,
in which case the one-component ($N_\uparrow=0$ or $N_\uparrow=N$)
boson and fermion systems have different spectra. In this case the $I_j$:s are integers
(half-odd integers), and the $J_\alpha$:s are half-odd integers (integers) if $N_\uparrow$
is even (odd). We note that $p$ will always be an integer. Thus, in practice,
all eigenvalues for $0<N_{\uparrow}<N$ can be found by 
letting $p$ run over all integers and choosing
all possible sets of $I_J$:s with the restriction $I_{\rm max}-I_{\rm min}<L$.

The total angular momentum depends on the above quantum numbers and
can be determined as 
\be 
M=\sum_{j=1}^N I_j +\sum_{\alpha=1}^{N_\uparrow} J_\alpha.
\ee

For  $N_\uparrow=0$ the quantum numbers $I_j$ are integers and $p=0$.
In this case we see immediately that the many-body energy (\ref{betheenergy})
 is a sum of the single-particle energies (\ref{tbse}). When $L\rightarrow\infty$ 
we recover the result for particles in a continuous ring and, say for 8 particles,
the spectrum shown with large blue dots in Fig. \ref{ener8}.
Next, consider the case $N_\uparrow=1$.  The number $p$ already goes through
all integers and, as predicted in the previous subsection, 
we thus get all possible energies, shown in small blue dots in 
Fig. \ref{ener8}.  Increasing  $N_\uparrow$ from one
thus can not give any new energy levels. A similar analysis for odd $N$ 
also shows that with $N_\uparrow=1$ we already get all possible energy levels.

We have here considered fermions. In the case of bosons the Hubbard model 
has turned out to be more difficult to solve with the Bethe ansatz method\cite{krauth1991}.
Fortunately, when we are only interested in the energy spectrum of a 1D ring
with infinitely strong  contact interaction, we do not need to solve the boson problem,
since the fermion solution gives the same energies as long as $0<N_\uparrow<N$,
as discussed in the previous section. 

\subsection{Noninteracting particles}

The single-particle spectra of noninteracting particles  
were given by Eqs. (\ref{spenwf}) and  (\ref{tbse}) for a 1D continuous ring and a TB lattice ring, respectively.
In the case of bosons the number of components (number of spin components)
does not change the energy spectrum, since each single-particle state may be occupied
by any number of bosons even in the single component case; only the degeneracies of the 
energy levels depend on the number of components.
At zero angular momentum all the bosons occupy the $\ell=0$ single-particle state. When the angular momentum is increased, the angular momentum 
state $\ell=1$ starts to be occupied until all particles are in that single-particle state.
Then we start to occupy the $\ell=2$ state and so on. 
The yrast energy increases linearly with the total angular momentum $M$, but
the slope has discontinuities at points $M=nN$, where the next single-particle
angular momentum starts to be occupied. Figure \ref{bf4} shows the beginning
of the yrast spectrum for eight noninteracting bosons (as pink dots).

\begin{figure}[h!]
\includegraphics[width=0.8\columnwidth]{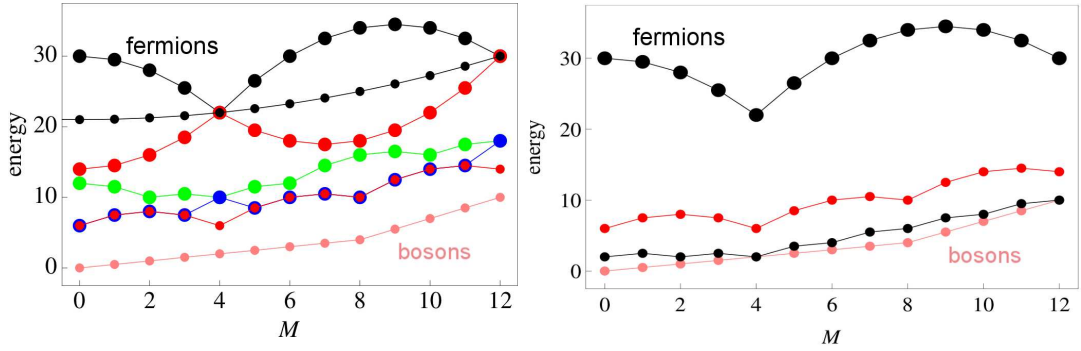}
\caption{
{\it Left:} Yrast spectra of eight noninteracting particles in a two-component system.
The lowest  (pink) curve shows the results for bosons: It only depends on the total number of
particles ($N_{\uparrow} + N_{\downarrow}$). The black curve shows the result for one-component fermions. 
The upper red curve represents $N_{\uparrow}=1$, the green curve corresponds to $N_{\uparrow}=2$, 
the blue curve to $N_{\uparrow}=3$, and the lower red curve (small dots) to $N_{\uparrow}=4$
(note that the blue and red curve coincide except at two points). The small 
black dots show the result for bosons and fermions with infinitely strong
contact
interaction for $0<N_{\uparrow}<8$. {\it Right:} Dependence of the yrast spectrum of noninteracting fermions
on the number of components (spin degrees of freedom). The one-component system is shown with
big black dots,  the two-component system with red dots, the four-component system with small black dots, 
and the eight-component system with pink dots.
Note that in the case of bosons the result (pink dots) is independent of the 
number of the components.
}
\label{bf4}
\end{figure}

In the case of fermions the number of components has a marked effect on the
spectrum of noninteracting particles. It is easy to construct the spectrum 
using the Pauli exclusion principle: 
Each single-particle state can accommodate only one (or zero) particle of each component.
Figure \ref{bf4} shows the yrast lines for 8 particles. The left panel shows the
result for a two-component system (spin-1/2 fermions) with different numbers of
spin-up and spin-down particles. The lowest energies are obtained when the number of 
spin-up and spin-down particles is the same. The left panel shows the effect of the 
number of components on the yrast spectrum of fermions. When the number
of components (spin of the particles) increases, the energy approaches 
that of bosons.  When the number of components equals that
of the number of particles, the fermion spectrum is identical to that of the bosons,
since in this case $N$ fermions can occupy the same single-particle state.

\section{Interaction with finite strength}
\label{finiteu}

So far we have only considered the ideal limits of zero and infinite (contact) 
interaction.
The more general case where the interaction strength is finite is naturally
more complicated. Nevertheless, also in this case several exact solutions exist.
The seminal paper for bosons was that of  Lieb and Liniger\cite{lieb1963} who
solved exactly the problem of interacting bosons in 1D, and for fermions
that of Lieb and Wu\cite{lieb1968} who solved the 1D Hubbard model for 
spin-1/2 fermions. Instead of reviewing this important analytic work, we
here choose to merely illustrate the qualitative differences between bosons and fermions, based on
exact (numerical) solutions of small Hubbard systems.

\subsection{Bosons with finite interaction}

In previous sections we have encountered examples of energy spectra for  
quantum rings of bosons with both infinite (Fig.\ref{ener8}) and zero interaction strength 
(Fig.\ref{bf4}). The intermediate problem of bosons interacting 
with finite contact interaction was solved exactly by 
Lieb and Liniger\cite{lieb1963}. Naturally, when the contact interaction increases 
from zero, the corresponding yrast lines smoothly interpolate between those of zero and 
infinitely strong interaction. 

\begin{figure}[h!]
\includegraphics[width=.5\columnwidth]{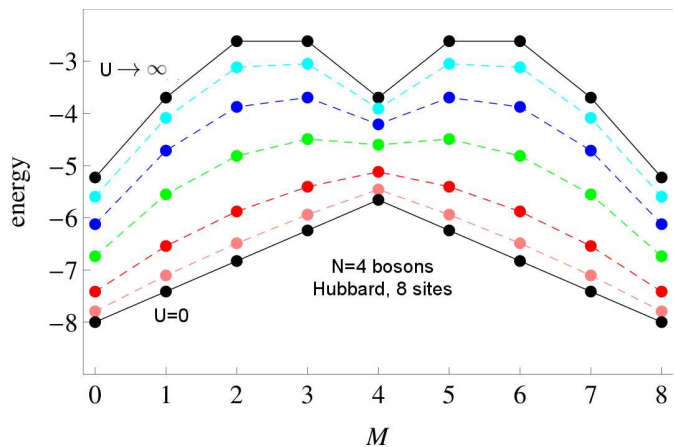}
\caption{Dependence of the yrast spectrum on the interaction strength $U$
of the Hubbard model for one-component bosons. The ring has 8 sites and 4 bosons.
The different curves represent different strengths of the interaction $U$. From bottom to
top $U=$  0, 0.3, 1, 3, 7, 20, and $U\rightarrow \infty$. The results for
$U=0$ and $U\rightarrow \infty$ are 
joined with solid lines, the others with dashed lines.
}
\label{bosonsU}
\end{figure}

To illustrate this, consider a small Hubbard ring with eight sites ($L=8$) 
and four particles ($N=4$). 
Figure \ref{bosonsU} shows the yrast energy as a function of the angular momentum
for different values of the interparticle interaction $U$ of the Hubbard model. 
The energy is a periodic function of the angular momentum $M$ due to
the finite number of sites. Nevertheless, even in this small (lattice) ring we already see 
clearly the same features as for a smooth 1D ring: In the case of noninteracting 
bosons the energy increases linearly with the angular momentum, and in the
case of $U\rightarrow \infty$ the energy has minima at $M=0$ and at $M=N$.
When $U$ increases from zero, the yrast energy smoothly approaches that
for $U\rightarrow \infty$. The local minimum at $M=4$ is reached with rather small interaction
strength $U=3$.

In the case of one-component (spinless) fermions the strength of the contact interaction
is irrelevant, since the Pauli exclusion principle prevents the 
particles from "seeing" the interaction.

\subsection{Yrast-line of two-component bosons and fermions}

Two-component systems have an effective spin 1/2. In the case of 
fermions it can be the true spin of the particles, while in the case of bosons
it can be the pseudospin, realized for example by two different hyperfine states of bosonic atoms.
While a contact interaction has no effect for {\it spinless} fermions, 
it does play a role when both spin components are present,
since the Pauli principle does not prevent particles with opposite spin 
from occupying the same site.
We learned above that in the case of an infinite interaction the 
energy spectrum is the same for bosons and fermions as soon as
we have at least one particle of the minority component.
This not true for interactions of finite strength. 

\begin{figure}[h!]
\includegraphics[width=0.8\columnwidth]{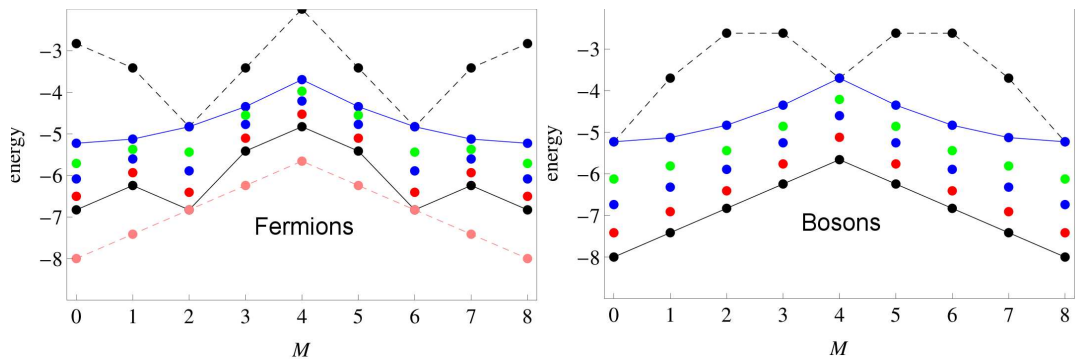}
\caption{Dependence of the yrast spectrum on the interaction strength $U$
of the Hubbard model. The ring has 8 sites and 4 particles, 2 spin-up and 2 spin-down.
The left panel shows the results for fermions and the right panel for bosons.
The black points connected with solid lines show the results for the noninteracting case
($U=0$), the blue points connected with solid lines correspond to
$U\rightarrow \infty$. 
Note that this result is the same for bosons and fermions for reasons discussed
in Section \ref{secinfu}. The upper curves with dashed lines show results for one-component 
systems. The boson result for the noninteracting case is plotted also in the left panel with pink points 
connected with dashed lines. The other plots in the fermion case correspond to $U=1$ (red),
$U=3$ (blue), $U=7$ (green), and in the boson case $U=0.3$ (pink),
$U=1$ (red), $U=3$ (blue).
}
\label{u422}
\end{figure}

We will first consider the yrast spectrum for a Hubbard system with eight sites and 
four particles so that we have two particles of each component.
Figure \ref{u422} shows the results for several strengths of the interaction $U$.
In the case of $U\rightarrow \infty$ the result is the same for fermions and bosons for
reasons described in Section \ref{secinfu}. In the case of noninteracting systems
the fermion and boson results differ, except at $M=2$ which is the ground state
for noninteracting fermions. When the interaction strength increases the yrast lines
smoothly  approach that of infinite $U$. In the case of fermions the local minimum
at $M=2$ remains to quite high values of $U$, it is still marked at $U=7$.
Only at very large $U$ the boson and fermion results will be qualitatively similar.

\subsection{Spin-splitting of the yrast line}

In the extreme case of infinite $U$, the effective Hubbard Hamiltonian is totally spin-independent, and thus the spectra are degenerate in the spin degree of freedom. As we shall see in this section, this is no longer the case for finite interaction strength $U$, where this degeneracy is split. First, however, let us return to the spin-independent limit for a two-component system, {\it i.e.} particles of (pseudo)spin 1/2. Different choices of numbers of particles in the up- and down components thus correspond to different $z$-components of the spin. 
For each total spin there will be a multiplet of
states, with the same energy and orbital angular momentum,  
corresponding to different $z$-components of the total spin or, equivalently,
different numbers of particles in each component.
In the case of four particles, for example, the energy spectrum computed
for two spin-up and two spin-down particles ($N_\uparrow=2$, $N_\downarrow=2$)
also includes all energy levels
for  $N_\uparrow=3$, and $N_\downarrow=1$ as well as 
for $N_\uparrow=4$ and  $N_\downarrow=0$.

\begin{figure}[h!]
\includegraphics[width=.45\columnwidth]{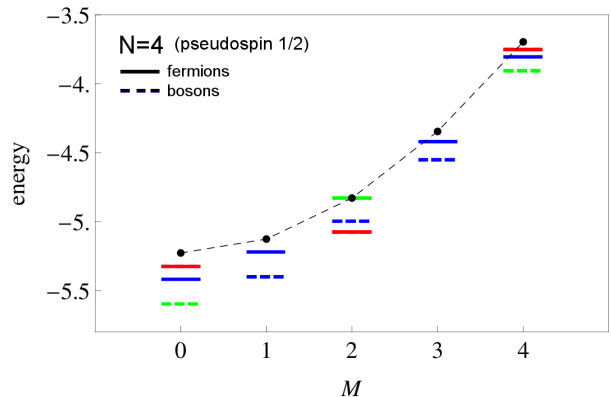}
\caption{Spin-dependence of the yrast spectra for four spin-1/2 fermions (solid lines)
and (pseudo)spin-1/2 bosons (dashed lines). The results are given for the Hubbard model with 8 sites
and $U=20$. Different colors indicate the total (pseudo)spin:
Red for $S=0$, blue for $S=1$ and green for $S=2$. The dashed and solid red lines
coincide. For infinite $U$ (dashed line with dots) the spin-splitting vanishes. For fermions the splitting at
large but finite $U$ corresponds to that of the antiferromagnetic Heisenberg model.
}
\label{spiny}
\end{figure}

The distribution ($N_\uparrow, N_\downarrow $) also provides information on 
the total spin of a given state of the two-component system.
In the case of four particles we know that the total spin can be $S=0$, 1, or 2. The totally polarized case
with $S_z=2$,
i.e. the one-component case, corresponds to $S=2$. Note that the case 
$N_\uparrow=3$, and $N_\downarrow=1$ has the $z$-component of
the spin $S_z=1$, but the total spin can be 1 or 2. Similarly, in the case
of  $N_\uparrow=2$ and  $N_\downarrow=2$, $S_z=0$, while the total spin can
be 0, 1 or 2.

In the case of an infinitely strong interaction, states with different total spin will
be degenerate. Away from this limit, however,  this is no longer true.
In particular, in the case 
of fermions the Hubbard model approaches the 
antiferromagnetic Heisenberg model for large but finite $U$ \cite{vollhardt1994},
so the Hamiltonian will contain a spin-dependent term
\be
H_{J} = \frac{J}{2} \sum_{i \neq j}^N {\bf S}_i \cdot {\bf S}_j.
\ee
Here, the effective Heisenberg coupling is proportional to the inverse of $U$ and vanishes
at $U\rightarrow\infty$
\footnote{It is well-known that the Hubbard model for fermions at {\it half-filling} and
in the limit of large $U$ approaches the antiferromagnetic Heisenberg 
model with effective coupling constant $4t^2/U$ \cite{vollhardt1994}. 
In fact, the large $U$ limit can be described by the 
Heisenberg model for {\it any} filling which is smaller than half-filling\cite{viefers2004},
though with a different coupling constant. }.
Thus, while the energies are independent of the total spin 
of the quantum state in the infinite $U$ limit, this degeneracy will split for finite $U$.
Note that this splitting is caused by the (spin-independent) 
inter-particle interaction $U$.

Figure \ref{spiny} shows the spin splitting of the lowest energy states for four 
particles in a Hubbard ring with eight sites. The dashed line with dots shows the 
result for $U\rightarrow \infty$. This result is the same for bosons and fermions and naturally does not
show any spin splitting. The colored lines show the energy levels for 
different spin states in the case when $U=20$. In the case of fermions (solid lines)
the spin splitting corresponds to that of the antiferromagnetic Heisenberg model. 
The lowest state for $M = 0$ has $S  = 1$. This is a result of Hund's
first rule: The lowest single-particle state ($\ell= 0$) is filled with two particles with opposite
spins, while the
two remaining particles occupy the states $\ell = 1$ and $\ell =  -1$ and
prefer to have the same
spin for removing the effect of the repulsive interparticle interaction.
The energy of the $S=2$ state for fermions equals that for infinitely strong interaction.
This is because now the system is polarized and thus a one-component system where
the interaction does not have any meaning.

The result for bosons is qualitatively different. Now the lowest state has $S=2$,
corresponding to a one-component system. The results for $S=0$ are 
the same as for fermions. This is true for any value of $U$ as seen also in
Fig. \ref{u422} where all the points for $M=2$ are the same for bosons and fermions.
In the case of four particles this result appears to be independent of the length
of the ring. However, computations for six particles show that this is not a general
property of bosons and fermions in Hubbard rings.

A major difference between the fermionic and bosonic cases lies in their respective magnetic states.
It has been shown that the Bose-Hubbard model for spin-1/2 particles 
with large on-site repulsion can be approximated by a {\it ferromagnetic} 
Heisenberg model\cite{zvonarev2009}.  Again for
half-filling the coupling constant has the same
value as in the case of fermions ($J=-4t^2/U$) but now the sign is
opposite.
In the case of four particles in a ring, the Heisenberg model (independent
of the sign of $J$)
is exactly solvable (see an exercise in[60]).
We notice that the splitting seen in Fig. 9 is already qualitatively the
same as in the Heisenberg model,
and that the boson and fermion cases have different magnetic order.
Numerical diagonalization of the Hubbard and Heisenberg Hamiltonians for
four and six fermions and bosons shows that
the coupling constant for large $U$  is the same for the antiferromagnetic
fermion case and the ferromagnetic boson case
for all filling fractions ($N \le L$). For half-filling ($N/L=1$),
$|J|\approx 4t^2/U$, while for smaller fillings $|J|$ decreases with $N/L$[4].

\subsection{Fixed points in fermion spectra}

Considering particles with infinitely strong contact interactions
we learned that in a two-component system it is sufficient to have only
one particle of the minority component in order to obtain the entire rotational spectrum
for any mixtures,
while the degeneracies will depend on the numbers of particles
in each component. 
In the case of one-component (spinless)
fermions the contact interaction does not have any effect and is equivalent
to the noninteracting case. At angular momenta $M=N/2+kN$ for even number of 
fermions and at $M=kN$ for odd number of fermions ($k$ any integer),
the lowest energy is the same for spinless fermions and for
 two-component systems with $N_{\downarrow}=1$ interacting with contact 
interactions.
 Moreover, it is easy to discover that the same energy results also for 
 noninteracting two-component  fermions with $N_{\downarrow}=1$, as seen
 in Fig. \ref{bf4}.
 
\begin{figure}[h!]
\includegraphics[width=0.8\columnwidth]{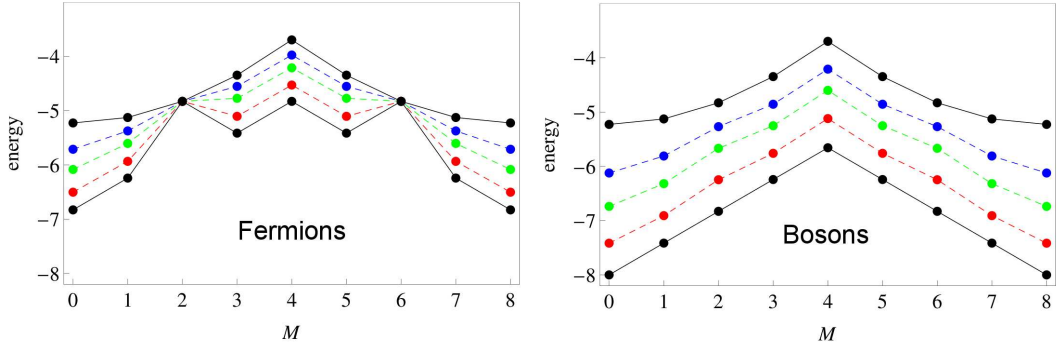}
\caption{Dependence of the yrast spectrum on the interaction strength $U$
of the Hubbard model in the case with $N=4$, $N_\uparrow=3$, $N_\downarrow=1$
and $L=8$. 
The left panel shows the results for fermions, and the right panel for bosons.
The curves from down to up correspond to
$U=0$, 1, 3, 7, and $U\rightarrow \infty$, respectively. Note that in the case of fermions the energy
at $M=2$ and $M=6$ is independent of $U$.
}
\label{fixedp}
\end{figure}

 This gives us the interesting result that at these angular momenta
 ($M=N/2+kN$ or $M=kN$),
 as long as $N_{\downarrow}=1$, the lowest  energy is completely independent 
 of the strength of the contact interaction. Naturally, this result only holds for fermions
 as demonstrated in Fig. \ref{fixedp}, showing the results for 
$N=4$, $L=8$ for bosons and fermions (now $N_\uparrow=3$ 
 and $N_\downarrow=1$).
 
\section{Quasi-one-dimensional rings}
\label{quasione}

In experiments, as they were mentioned in the Introduction,
the quantum rings will typically not be strictly one-dimensional, which 
means that the particles eventually may ``pass'' each other. 
Considering the ring as a quasi-one-dimensional (Q1D)
wire, the perpendicular modes of the single-particle wave function
can be separated from the longitudinal modes. If the excitation energy 
of the perpendicular modes is large, the many-particle state is composed 
mainly of the lowest perpendicular mode, and the system becomes rather similar
to the strictly 1D case.

\subsection{Continuous Q1D rings}

A generic continuous Q1D ring can be modelled using a harmonic confinement,
\be
V(r,z)=\frac{1}{2}m\omega^2(r-R)^2 + V_z
\label{annpot}
\ee
where $R$ is the radius of the ring, $r$ the radial coordinate on the plane of the ring, 
$\omega$ the planar confinement frequency and $m$ the mass of the particles. 
$V_z$ indicates the confining potential perpendicular to the plane. 
Depending on 
the system considered, it might be harmonic or, {\it e.g.},
correspond to the effective confinement experienced by the 2D electron gas in semiconductor heterojunctions.
Frequently $V_z$ is assumed to be so large that motion in the $z$-direction is completely frozen out, 
and thus the system is effectively two-dimensional.
We will here only consider 2D systems and thus the second term
in Eq. (\ref{annpot}) is not needed.

\begin{figure}[h!]
\includegraphics[width=.6\columnwidth]{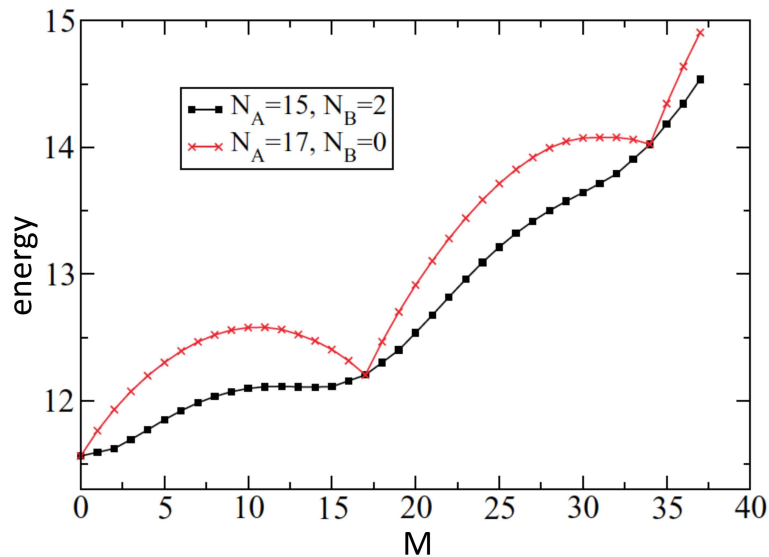}
\caption{The yrast energy as a function of the angular momentum for a 
Q1D ring with 17 bosons. The red line is for a one-component system and
the black curve for a two-component system with 15 particles in the majority component and 
2 particles in the minority component. The energy is in units of the effective oscillator 
strength of the confinement. From~\cite{bargi2010}.
}
\label{bargi1}
\end{figure}

Exact diagonalization techniques as well as density functional methods
have been used to describe electrons in Q1D rings\cite{chakraborty1994,koskinen2001}. 
For reviews we refer to\cite{viefers2004,reimann2002}. 
Exact diagonalization studies show that in narrow rings the spin and charge 
excitations separate: The former resemble those of 
an antiferromagnetic Heisenberg model, and the latter vibrational modes
of localized electrons\cite{koskinen2001,viefers2004}.
The results of density functional methods (making use of the local-spin-density 
approximation) show clearly the electron localization in narrow rings and
give qualitatively correct persistent 
currents\cite{reimann1999,emperador1999,viefers2000,emperador2001}.

Bosons in annular traps have been studied using the Gross-Pitaevskii
method(see for example, \cite{cozzini2006,abad2008,ogren2009,malet2010,abad2010,abad2011,adhikari2012}) as well as with exact diagonalization 
(see for
example,\cite{jackson2006,bao2007,smyrnakis2009,mason2009,karkkainen2007,bargi2010,kaminishi2011}).

Figure \ref{bargi1} shows the yrast spectrum for 17 bosons in a Q1D quantum
ring, calculated by Bargi {\it et al.}~\cite{bargi2010}
using exact diagonalization techniques. In the case of a one-component
system (red line) the result is qualitatively similar to  that of particles interacting 
with an infinitely strong contact interaction (see Fig. \ref{ener8}).
In the two-component case, with $N_\uparrow=15$ and $N_\downarrow=2$,
a smoother curve is produced. In an infinitely narrow ring the black curve would 
be expected to be a parabola, for the reasons explained in Section \ref{secinfu},
independent of the number of particles in the minority component $N_\downarrow$
(as long as it is nonzero). It is interesting to note that the effect of the 
finite width of the confinement appears to be (qualitatively) stronger in the case
of a two-component system than in the case of a one-component system.

\subsection{Q1D rings in the Hubbard model}

\begin{figure}[h!]
\includegraphics[width=0.8\columnwidth]{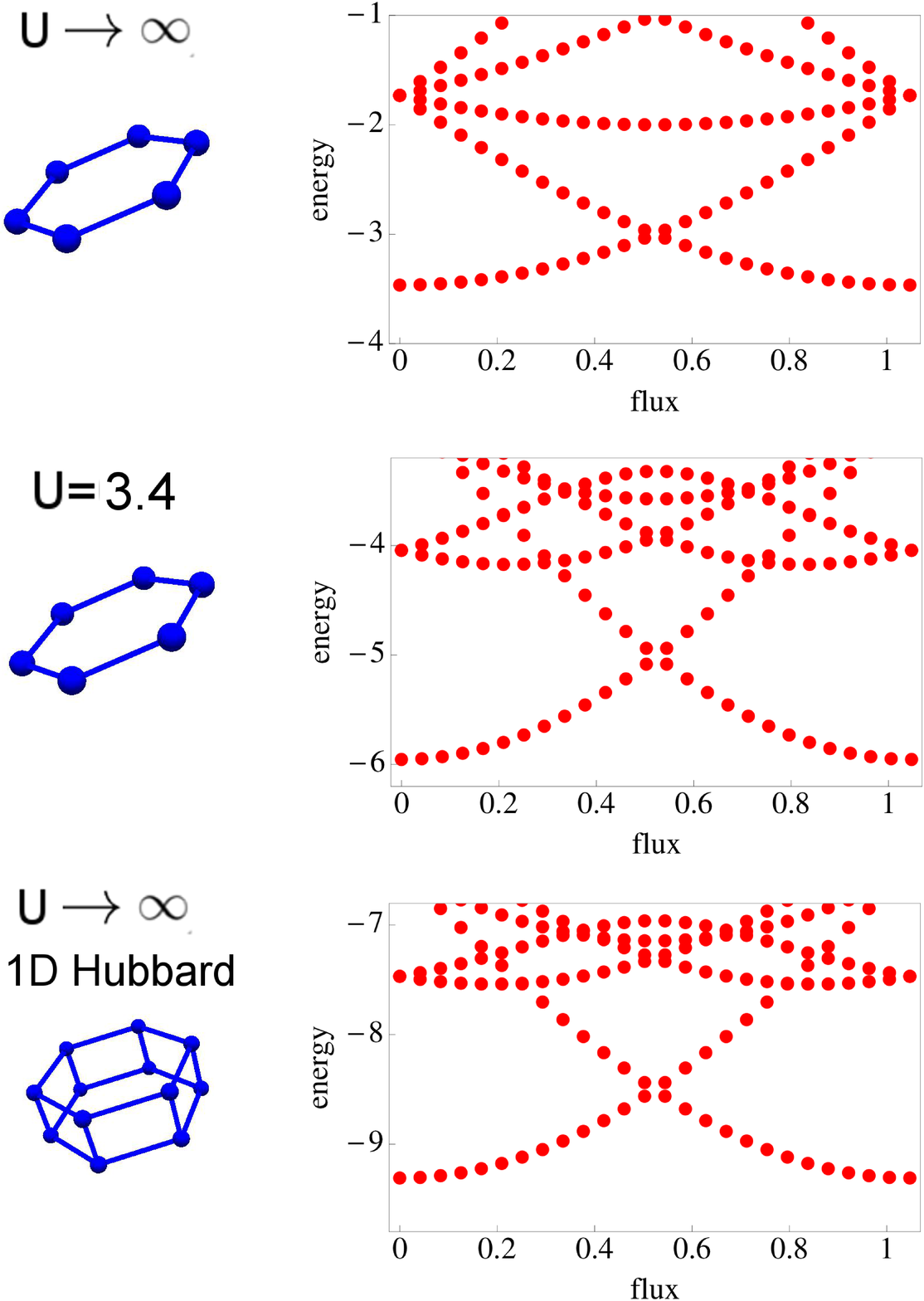}
\caption{Flux dependence of the lowest energy levels for four bosons
in quantum rings.
The uppermost panel shows the result for $U\rightarrow \infty$.
The middle panel shows the result for $U=3.4$. the lowest
panel shows the result for $U\rightarrow \infty$ in the case of a Q1D Hubbard ring.
The rings in each case are shown at the left.}
\label{q1d1}
\end{figure}

Quasi-one-dimensional quantum rings can also be made out of optical lattices 
which can be described with the Hubbard model. In the case of bosons the
simple 1D Hubbard ring with finite $U$ already mimics the Q1D ring,
since a finite $U$ means that the particles can pass each other like in a Q1D
continuous ring.
This is illustrated in Fig. \ref{q1d1}:  The 1D ring with a finite $U$ shown in the
middle panel of the figure, gives a qualitatively similar low energy spectrum 
as a Q1D ring with infinite $U$, shown in the lowest panel. 

\begin{figure}[h!]
\includegraphics[width=.5\columnwidth]{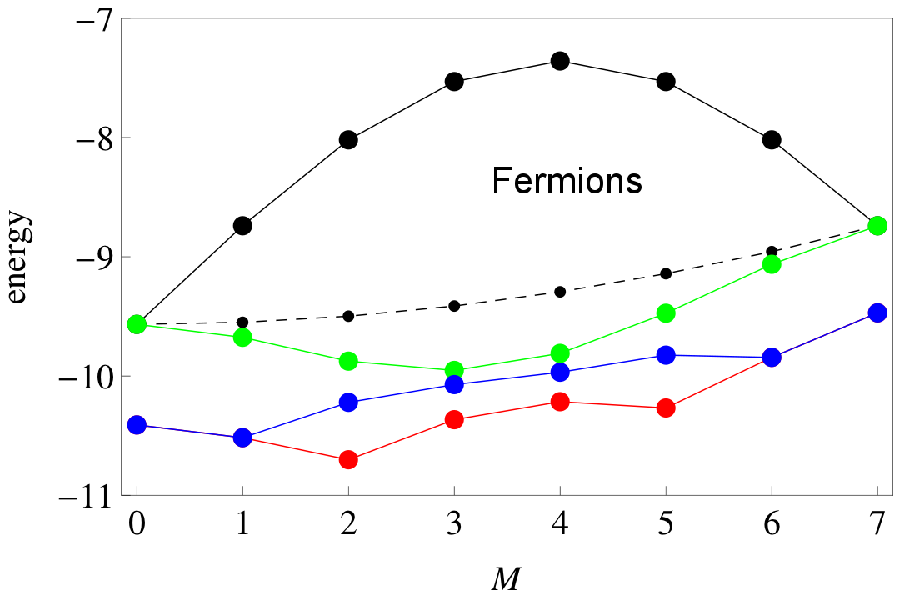}
\caption{Energy as a function of angular momentum for two-component 
fermions in a Hubbard ring with $L=15$ sites, $N_\uparrow+N_\downarrow=7$ 
particles and the interaction strength $U=5$. 
$N_\uparrow=7$ black; $N_\uparrow=6$ green; $N_\uparrow=5$ blue; $N_\uparrow=4$  red.
For comparison, the dashed line shows the $U\rightarrow \infty$ result which the same for $0<N_\uparrow<7$.
Note the qualitative similarity of blue curve ($N_\uparrow=N-2=5$) with 
the black curve in Fig. \ref{bargi1}.
}
\label{ef15}
\end{figure}

In the case of fermions the situation is different. In a strictly 1D one-component system the Pauli 
exclusion principle prevents particles from passing each other even with zero $U$.
However, in the case of two-component fermions the finite $U$ Hubbard model also mimics a 
Q1D ring\cite{viefers2004}. This is illustrated in Fig. \ref{ef15} which 
shows results for seven two-component fermions in a ring with 15 sites
with the interaction strength $U=5$. We know from the previous sections that
for $U\rightarrow \infty$ the yrast line is the same for any $0<N_\uparrow<N$. 
Figure \ref{ef15} shows that this is not the case for finite $U$, but results for different
$N_\uparrow$ differ qualitatively. It is interesting to note that the case 
$N_\uparrow=N-2$ is qualitatively similar to the related case for bosons in a
Q1D ring, shown in Fig. \ref{bargi1}.

\begin{figure}[h!]
\includegraphics[width=0.8\columnwidth]{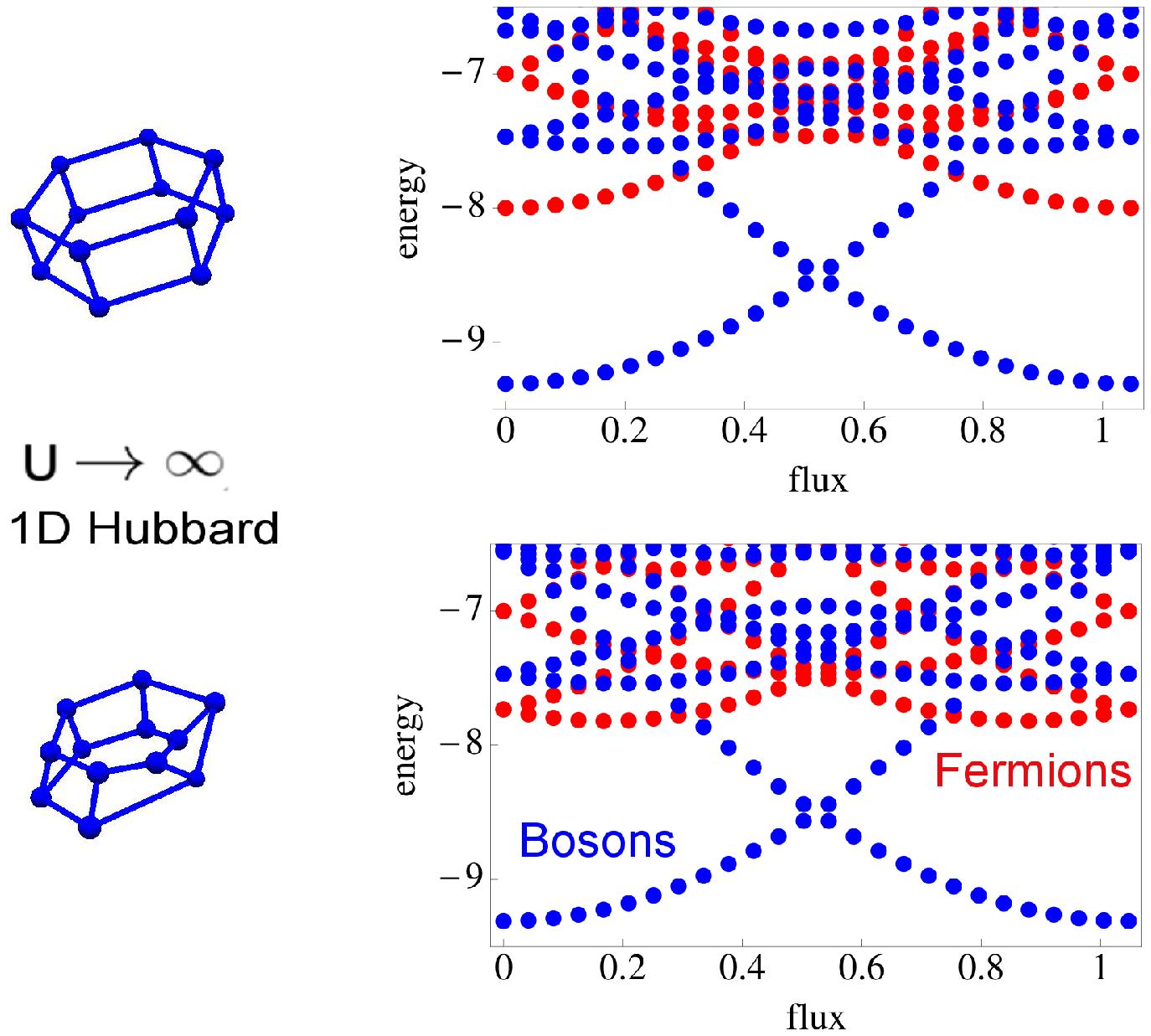}
\caption{Flux dependence of the lowest energy levels for one-component fermions (red)
and bosons (blue) interacting with infinitely strong contact interactions
for the different Hubbard rings shown left. In each case the system has four particles.
The lower case is the M\"obius ring.
}
\label{q1d2}
\end{figure}

The finite width of the quantum ring can change the energy spectrum markedly.
In the case of four fermions in a 1D ring the ground state is degenerate and has 
a finite angular momentum. This is not necessarily the case if the ring is wide enough.
The upper panel of Figure \ref{q1d2} shows that in a Q1D Hubbard ring the
lowest energy state of the fermion spectrum is not degenerate and the energy 
increases with the flux, even if the system has four electrons. In the 1D case
the ground state would be degenerate at zero flux, and increasing the flux would
split the degeneracy, decreasing one of the energy levels.

Figure \ref{q1d2} also shows the result for a so-called M\"obius ring, studied in detail by
Ballon {\it et al.}\cite{ballon2008}, where the fermion ground state is degenerate.
It is interesting to note that in the M\"obius ring the boson spectrum is nearly
identical with that of the normal Q1D ring, while the fermion spectrum 
becomes qualitatively different. Note that only the periodic boundary condition is changed
in going from the normal ring (upper panel in the figure) to 
the M\"obius ring (lower panel).
 
 \section{Conclusions}
 The theoretical description of quantum rings is a vast field, involving various types and strengths of interactions, continuum- versus lattice models,
 strictly 1D versus quasi-1D systems, presence or absence of magnetic fields
 (and thus persistent currents), spinful versus spinless particles -- and of
 course the quantum statistics of the particles. Our hope is that the present
 paper has provided a comprehensive and fairly self-consistent overview of the
 similarities and differences between bosonic and fermionic quantum rings in
 all of the above contexts. As we have seen, in many cases the quantum
 statistics of the particles does make a big difference to the energy spectra,
 and thus the physical properties, of the system. Since both bosonic and fermionic quantum rings can be fabricated in the lab nowadays, these differences are important to be aware of. 
 The list of references certainly does not exhaust all of the work done in the field, in particular since we have chosen to mainly focus on small systems. 
 
 \eject

\acknowledgments

We thank G. Kavoulakis, F. Malet and E. \"Oznur Karabulut for useful
discussions. This work was financially supported by Academy of Finland, the Norwegian Research Council, the Swedish Research
Council, and the Nanometer Structure Consortium at Lund University. 

\bibliography{references_arxiv}

\end{document}